\documentclass[12pt]{article}

\usepackage{times}

\usepackage{graphicx}
\usepackage{caption}

\usepackage[numbers,sort&compress]{natbib}

\newcounter{lastnote}

\title{Response to Comment (arXiv:1506.02787v1) on Selective Interface Control of Order Parameters in
Complex Oxides }

\author
{D. Meyers$^{1,\ast}$, Jian Liu$^{2,\ast}$, J. W. Freeland$^{3}$, S. Middey$^{1}$, M. Kareev$^{1}$, J. M. Zuo$^4$, \\ Yi-De Chuang$^5$, Jong Woo Kim$^{3}$, P. Ryan$^{3}$, and J. Chakhalian$^1$\\
\\
\normalsize{$^{1}$Department of Physics, University of Arkansas, Fayetteville, AR 72701, USA}\\
\normalsize{$^{2}$Department of Physics, University of California, Berkeley, CA 94720, USA}\\
\normalsize{$^{3}$Advanced Photon Source, Argonne National Laboratory, Argonne, IL 60439, USA}\\
\normalsize{$^{4}$Department of Materials Science and Engineering, University of Illinois, Urbana, IL 61801, USA}\\
\normalsize{$^{5}$Advanced Light Source, Lawrence Berkeley National Laboratory,
Berkeley, CA 94720, USA}\\
\\
\normalsize{$^\ast$ Both authors contributed equally; E-mail: dmeyers@uark.edu, jian.liu@berkeley.edu}
}

\date{}
\usepackage{color}

\begin{document} 

\baselineskip24pt
\maketitle 

{\bf In response to  Lu \textit{et  al}, (arXiv:1506.02787v1), here we present a detailed writeup concerning the questions raised in their comment on our eprint (arXiv:1505.07451). The key question raised by Lu \textit{et al} was if the bulk-like charge ordered state becomes indetectable with resonant scattering due to ultrathin film thickness. In this reply, we first detail the relation of our work to past work on the same compound by Staub \textit{et al} to demonstrate that the presented data are indeed sufficient to support our claims of no charge order on ultra thin films of NdNiO$_3$ (NNO) on NdGaO$_3$ (NGO). Further, we demonstrate that if a well defined charge ordered phase exists in ultra thin films, it is indeed resolvable such as that in EuNiO$_3$ (ENO).}

Regarding the comments by Y. Lu, E. Beckiser, and B. Keimer on our manuscript\cite{ArxivDM}, here we offer our response to their concerns. The  questions raised concerned the sensitivity of resonant scattering to show the absence of a symmetry change in our films across the metal-to-insulator transition (MIT). It is first helpful to review key findings and results by resonant scattering in the literature. In the work of Staub et al.\cite{Staub} on thick films ($\sim$ 130 unit cells (u.c.) of NNO), the experiment tracked two aspects of the scattering as direct evidence of charge ordering. The first aspect was concerning the fact that across the orthorhombic to monoclinic phase transition, a (015) peak, which is forbidden in the orthorhombic phases, appears at the MIT due to the lowering of symmetry to a monoclinic space group with two non-equivalent Ni sites. In addition, both the (105) and (015) structure factors develop a contribution from the different nature of the two Ni sites, which is not present in the (105) peak in the orthorhombic phase. For the thick films, this measurement leads to  a clear Ni K-edge resonant signature on both peaks that collapsed to zero at the transition to the high-temperature orthorhombic phase, with bulk-samples showing similar results\cite{Staub,Lorenzo}. Moreover, this resonant effect has opposite signs between the two peaks. The key finding of our eprint on ultra-thin NdNiO$_3$ films is that there is no resonant signature on either peak  below the MIT and the (015) peak is absent at all temperatures.

In the comment\cite{Comment}, Lu et al. argued that the observation of the Ni resonance on the (220) peak does not mean the Ni resonance on the (105) reflection is detectable because the (220) has a much larger structure factor than the (105), and while the (220) resonance  is proportional to the addition of the  Ni form factors of the two different sites, the (105) reflection involves a subtraction of the same. First of all, we would like to clarify  that we used the measurement of the (220) peak only as a definitive marker of the resonant Ni K-edge energy position and did not seek to imply that the observation of the (220) reflection intensity guarantees that the (105) peak will always show energy dependence at the Ni K-edge with inequivalent Ni sites. In other words,  the (220)\ peak was instead added as a proof of concept to show that the beam line in this configuration is certainly capable of observing Ni resonance, i.e. to eliminate an experimental issue as the culprit for the lack of an observed resonance enhancement at the (105). Meanwhile, whereas  the comment argues the (220) reflection intensity is four orders of magnitude larger then the (105) intensity based upon bulk structure,  inspection of the experimental (220)/(105) ratio gives a value of $\sim$ 150, or two orders of magnitude\cite{ArxivDM,Munoz}. Nevertheless, as discussed below, the relative intensity of the (220) and the (105) peaks are not important to our conclusion, as we do observe the (105) peak well above the noise, but no resonance response indicative of a contribution from Ni.

Lu \textit{et al} also show that the intensity of the (105) peak is dependent on $B_{O,Nd}$, the allowed Thompson scattering term from the Nd and O sites, and the difference of the real and imaginary parts of the Ni site resonant correction term. They argue the relative weakness of the (105) reflection and the substrate background could complicate this measurement and prevent detection of the resonance signature. Staub \textit{et al} showed, on thick films grown on the same substrates (NGO) at  the (105) reflection, despite the relative weakness of this peak and the substrate background, that upon traversing the metal-insulator transition (MIT) the resonance contribution ($\sim$ 31\%) is significant, and a clear increase in the off-resonance scattering is also present\cite{Staub}. There is no reason to conclude that ultra thin geometry would selectively or significantly weaken the resonant contribution compared to the Thompson term if the same Ni-site charge order was present.  In the bulk-like film, the resonant enhancement leads to a $\sim$ 31\% intensity increase of the (105) peak at the Ni K-edge. For our measurement, we observe no resonance peak and noise of approximately 5\% of the total intensity, thus if charge order exists we can place an approximate upper limit on the charge disproportionation of 0.073$\pm$0.007e, which is significantly less than in the bulk (0.45$\pm$0.04e)\cite{Staub}. Further, the lack of any observable temperature dependence of the peak, in contrast to the peak intensity increases with reduced temperature found by Staub \textit{et al}, would reduce this upper limit even further\cite{Staub}. Thus, our combined results of  observing a \textit{clear (105) reflection with no temperature dependence} and \textit{no resonant contribution} provides strong evidence  for  our conclusion that the bulk-like CO transition is absent in the ultra-thin limit. This is our primary evidence presented in the eprint, as the observation of the scattering peak precludes the possibility of the bulk-like resonance enhancement ($\sim$ 31\%) being too weak to observe as per Staub \textit{et al}'s results, as potentially might be the case for the (015) reflection.

The second comment from Lu \textit{et al} concerns the (015) peak being significantly weaker then the (105) peak,  i.e. $\sim$ 1000 times when refinement of the bulk structure is considered\cite{Munoz}. They also acknowledge that Staub \textit{et al} found an experimental ratio to  be closer to $\sim$ 10 instead and argue that this large difference could be due to twinning in the films used by Staub \textit{et al \cite{Staub}}. We cannot  comment on   the potential existence of twinning in the materials structure used in \cite{Staub} since we were not involved in that study.  However, if Staub's film does contain twinning, this would imply that the resonance contribution to the (105) was reduced, as the (105) and (015) resonant contributions have opposite sign, thus making the resonance a more significant percentage of the total signal in a non-twinned sample. On the other hand, if the (105)/(015) ratio was not due to twinning on Staub's films, then the experimentally observed  (105)/(015) $\sim$ 10 ratio is expected, as shown in Fig. 1(a) (below). Therefore, if the same charge ordering is present in our sample, the (015) film peak would have been clearly detected, which is however not the case. Further, the actual structure factors in thin films will be different from bulk as the substrate effects the structure\cite{Ramesh,Ramesh2}. While Lu \textit{et al} are also concerned that the substrate reflection dominates at the H or K $= 1$ r.l.u. (reciprocal lattice unit) position, the ultra thin film geometry actually becomes very helpful here by broadening the film peak, which allows observation of the film peak in the vicinity of the very intense but also extremely sharp substrate peak. Thus, compared with the Staub \textit{et al.} experimental result, despite the fact that the structure factor is indeed  weaker, the lack of a (015) reflection is still consistent with our conclusion that the CO is absent below the MIT\cite{Staub}. 

Finally, as \textit{direct evidence} that one can observe charge-ordering even in ultrathin films of nickelates by  resonant X-ray  scattering, we include here our recent data on a  14 unit cell EuNiO$_3$ (ENO) grown on the same NGO  substrates with data taken under  similar   experimental conditions at the 6-ID-B beam line of the Advanced Photon Source. For the ENO case, the strong electron-phonon coupling due to the large lattice distortion from  the smaller rare earth cation makes it less likely to suppress the symmetry transition at the MIT. To demonstrate that the (0KL) reflections, with K and L odd, are observable even in 14 u.c.  ultra thin films, here we include a low temperature (below MIT), 100K, scan of the Ni K-edge resonance at the (011) reflection for a 14 unit cell ENO film grown on NGO, shown in Fig. 1(b) (below). As with the (015), and any (0KL) reflection, with K and L odd, this peak is forbidden in $Pbnm$ orthorhombic symmetry but becomes allowed for the Ni sites in the $P2_1/n$ monoclinic phase. As immediately  seen,  the ultra thin film geometry and close proximity to the substrate reflection \textit{does not preclude observation of this peak or a resonant enhancement}, as we observed a strong film reflection off-resonance and a $\sim$ 65 \% resonant enhancement at the Ni K-edge corresponding to  the non-equivalent Ni sites and symmetry  lowering below the MIT. However, it is still possible the (015) reflection of NNO was below the detection limit in the experiment, however the (015) data is presented only as additional supporting evidence to the (105) data that conclusively shows no resonance response across the MIT. Thus, these results further  corroborate the extreme sensitivity of resonant x-ray scattering to detect the charge ordered state in nickelates and  further evidences  our conclusion that the absence of  both the resonant enhancement and  temperature dependence of the (105) reflection and the lack of any (015) reflection in our NNO films clearly supports the e-print notion on the  suppression of the CO ground state while preserving a bulk-like metal-insulator transition and E$^{\prime}$-type anti-ferromagnetism.

In addition, we note that several relevant references were unintentionally  omitted in our eprint and will  be added to the coming revised version\cite{Hepting,Wu,Frano,Hwang}.

\begin{figure}
\centering
\includegraphics[width=.7\textwidth]{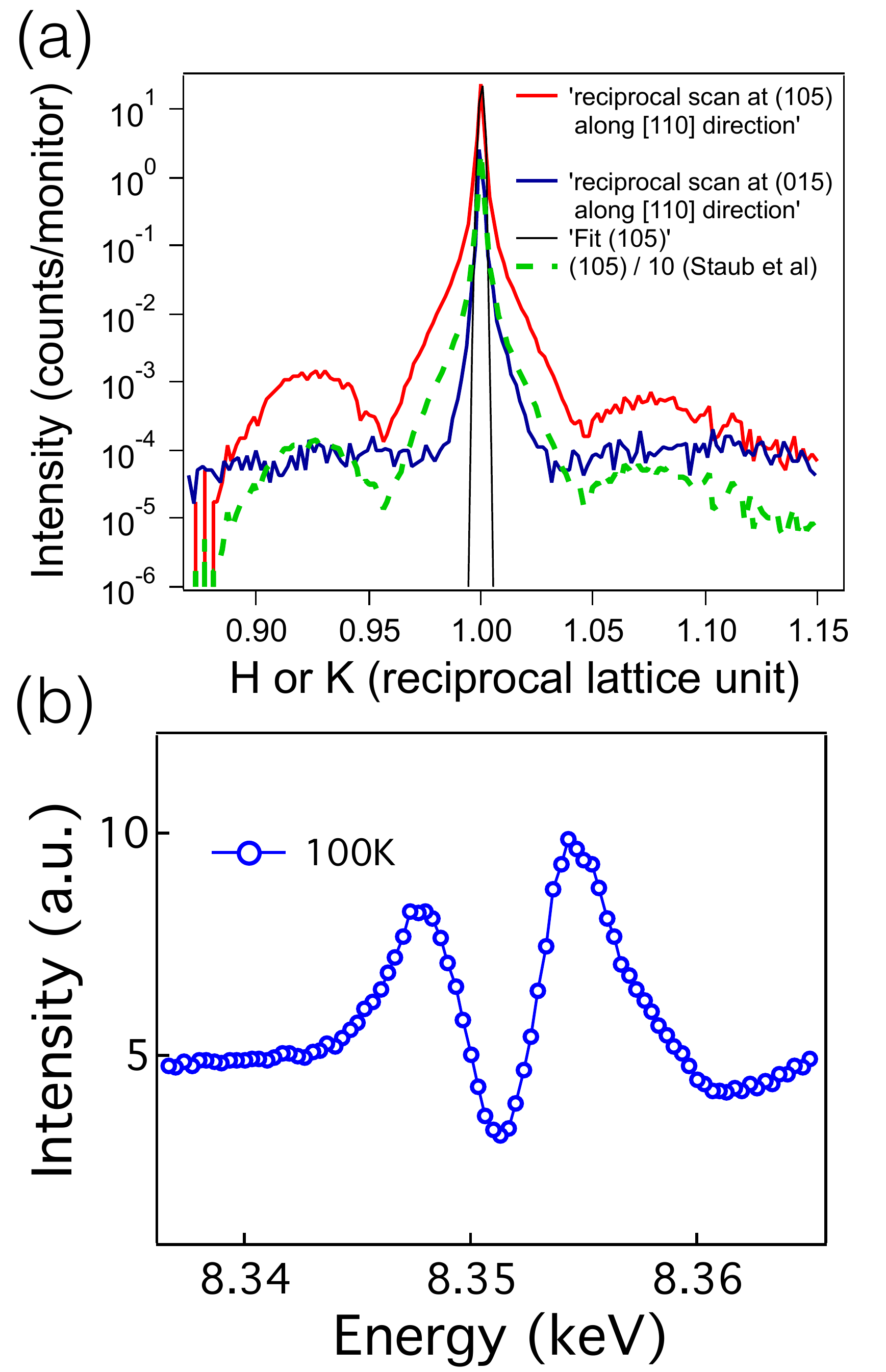}
\caption{(Color online) (a) The (105) and (015) measurements, compared to the (105) peak divided by ten. The film peak would still clearly be observable. (b) Ni K-edge scan of the (011) reflection on a 14 unit cell EuNiO$_3$ film on a NGO substrate, showing that the resonance response of the reflection is  observable in ultra thin films.}
\end{figure}

\end{document}